\documentclass[aps, prb, preprint  groupedaddress, notitlepage]{revtex4-1}
\usepackage{geometry}                
\usepackage{graphicx}
\usepackage{amssymb}
\usepackage{epstopdf}
\usepackage{slashed}
\usepackage{color}
\usepackage{amsmath}
\newcommand{\slp}{\slashed{p}}
\newcommand{\tr}{\text{tr}}

\newcommand{\cep}{\chi^{\text{e}+}}
\newcommand{\cem}{\chi^{\text{e}-}}
\newcommand{\cop}{\chi^{\text{o}+}}
\newcommand{\com}{\chi^{\text{o}-}}
\newcommand{\be}{\begin{equation}}
\newcommand{\ee}{\end{equation}}
\newcommand{\fL}{\mathcal{L}}
\newcommand{\fH}{\mathcal{H}}
\newcommand{\fN}{\mathcal{N}}

\newcommand{\fO}{\mathcal{O}}
\newcommand{\sld}{\slashed{\partial}}
\usepackage{soul}
\newcommand{\stp}{\text{\st{p}}}

\newcommand{\stk}{\text{\st{k}}}

\begin{document}

\title{Renormalization group analysis of phase transitions in the two dimensional Majorana-Hubbard model }
\author{Kyle Wamer and Ian Affleck}
\affiliation{ Department of Physics and Astronomy and Stewart Blusson Quantum Matter Institute, University of British Columbia, 
Vancouver, B.C., Canada, V6T1Z1}
\date{\today}                                           

\begin{abstract}

A lattice of interacting Majorana modes can occur in a superconducting film on a topological insulator in a magnetic field. The phase diagram as a function of interaction strength for the square lattice was 
analyzed recently using a combination of mean field theory and renormalization group methods, and was found to include second order phase transitions. One of these corresponds to spontaneous breaking of an emergent U(1) symmetry, for attractive interactions. Despite the fact that the U(1) symmetry is not exact, this transition was claimed to be in a supersymmetric universality class when time reversal symmetry is present and in the conventional XY universality class otherwise. Another second order transition was predicted for repulsive interactions with time reversal symmetry to be in the same universality class as the transition occurring in the Gross-Neveu model, despite the fact that the U(1) symmetry is not exact in the Majorana model. We analyze these phase transitions using a modified $\epsilon$-expansion, confirming the previous conclusions.

\end{abstract}

\maketitle

\section{Introduction}

\begin{figure}[ht]
\begin{center}
\includegraphics[width=0.999\textwidth]{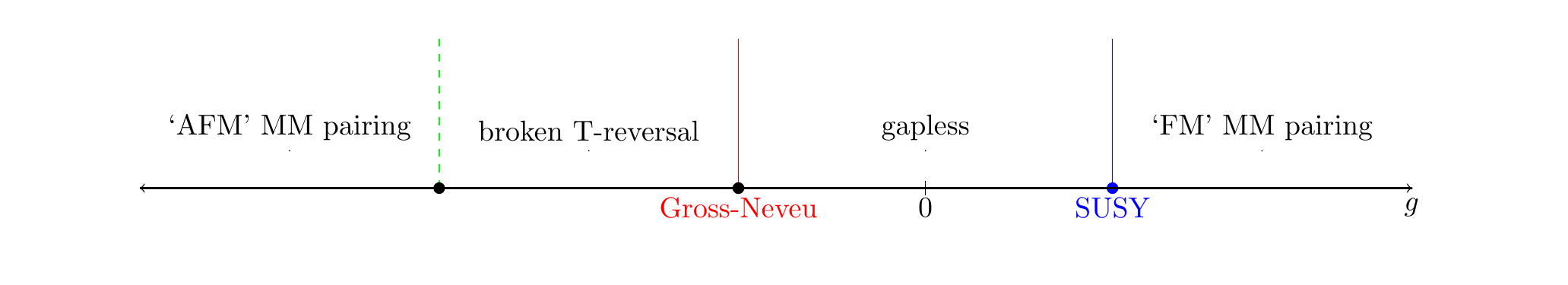}
\caption{Proposed phase diagram of the Majorana-Hubbard model with time reversal symmetry from [\onlinecite{PhysRevB.96.125121}]. The dotted line is a first order phase transition. `MM pairing' refers to the mean field prediction that Majorana modes will pair up on neighbouring sites to form Dirac fermions, breaking translation symmetry in either the horizontal or vertical direction. For $g>0$, the Dirac fermions' energy levels are empty (FM pairing), while they alternate being empty and occupied (AFM pairing) for $g<0$.}
\label{phasediagram}
\end{center}
\end{figure}

The observation of Majorana fermions in condensed matter has attracted great attention. A setting in which a macroscopic number of interacting Majorana fermions are predicted to occur is a layer of ordinary superconductor on a strong topological insulator in a transverse magnetic field.\cite{Alicea2012, Beenakker2013a} The resulting vortex lattice is predicted to have a Majorana mode localized at every vortex core.\cite{RevModPhys.87.137}  The interactions between the Majorana modes are predicted to drop off exponentially with the superconducting coherence length. \cite{Chiu2015} The simplest model for this system, the ``Majorana-Hubbard model'', has nearest neighbor hopping and the shortest possible range interaction, which must occur on 4 sites;\cite{2015PhRvL.115p6401R, 2015PhRvB92w5123R, Milsted2015, Chiu2015, PhysRevB.96.125121, Grosfeld2006} for the square lattice, these are plaquettes:
\be \label{1:1}
	H = it\sum_{m,n}\gamma_{m,n}[(-1)^n\gamma_{m+1,n} + \gamma_{m,n+1}] +  g\sum_{m,n}\gamma_{m,n}\gamma_{m+1,n}\gamma_{m+1,n+1}\gamma_{m,n+1}
\ee
A similar model, on the honeycomb lattice, was considered recently in [\onlinecite{2018arXiv180606092L}]. The operators $\gamma_{m,n}$ are Hermitian, and satisfy the anti-commutation relations
\be
	\{\gamma_{m,n},\gamma_{m',n'}\} = 2\delta_{m,m'}\delta_{n,n'}.
\ee
This model was studied in detail in [\onlinecite{PhysRevB.96.125121}], where it was shown to have a rich phase diagram as a function of $gt^{-1}$. The strong coupling limit was also studied recently in [\onlinecite{2017arXiv171103632K}]. In particular, second order phase transitions were identified at $g= g_{\text{c},1} \approx -0.9t$ and $g = g_{\text{c},2} \approx +0.9t$. The arguments of [\onlinecite{PhysRevB.96.125121}] were based on a combination of mean field theory and renormalization group methods, and involved finding the low energy continuum limit of (\ref{1:1}). By neglecting derivative corrections, the corresponding field theory was shown to have emergent Lorentz invariance and an emergent U(1) symmetry. In terms of a 2-component complex fermion $\psi$, the imaginary time Lagrangian density is
\be \label{1:2}
	\fL_1 = \bar\psi \gamma^\mu \partial_\mu \psi + 64g\Lambda_0^{-2}(\bar\psi\psi)^2.
\ee
Here $\bar\psi := \psi^\dag\gamma^0$, and the Eucliean Dirac gamma matrices, built out of Pauli matrices according to  $\gamma^\mu = \{\sigma_y, \sigma_x,-\sigma_z\}$, satisfy
\be \label{eq:euclidgamma}
	\{\gamma^\mu, \gamma^\nu\} = 2\text{diag}(1,1,1).
\ee
The coefficient $\Lambda_0=a^{-1}$ is a bare cutoff defined by the inverse  of the lattice spacing, $a$, and the time coordinate has been rescaled so that the velocity $v=4ta \equiv 1$. It was argued that the critical point $g_{\text{c},1}$ is in the universality class of the Gross-Neveu model, while the point $g_{\text{c},2}$ corresponds to the $\mathcal{N}=2$ supersymmetric universality class (see Figure \ref{phasediagram}).

In this paper, we consider the effects of adding U(1) and time reversal breaking operators to $\fL_1$, and ask if this changes the universality classes of the transitions at $g_{\text{c},1}$ and $g_{\text{c},2}$. In Section \ref{section:field}, we review and extend the low energy field theory describing the predicted gapless phases of this model, and determine the leading U(1) breaking operators. In Section \ref{section:modified}, we introduce a modified $\epsilon$-expansion which is able to treat three dimensional operators that break Lorentz invariance in four dimensions, without generating unphysical contributions. In Section \ref{chapter:u1}, we show that all U(1) breaking operators are irrelevant to one loop order, using this modified $\epsilon$-expansion and Wilsonian renormalization. In Section \ref{chapter:mass}, we consider the effects of a time reversal breaking perturbation, a fermion mass term, on the Majorana-Hubbard model. Using a combination of renormalization group and supersymmetry methods, we show that such a perturbation is relevant, and changes the universality class from the $\fN=2$ supersymmetric one to the conventional $XY$ transition. Section \ref{section:conclusion} contains our conclusions. 

\section{Low Energy Field Theory} \label{section:field}

Due to the alternating nature of the nearest neighbor hopping in (\ref{1:1}), the unit cell spans two lattice sites, so we define
\be \label{eq1:0}
	\gamma_{m,2n} = \gamma^{\text{e}}_{m,2n} \hspace{10mm} 
	\gamma_{m,2m+1} =\gamma^{\text{o}}_{m,2n+1}.
\ee 
These definitions of $\gamma^{\text{e}/\text{o}}$ are slightly different than those of [\onlinecite{PhysRevB.96.125121}], and are chosen to simplify the form of the U(1) breaking operators. To derive a low energy field theory, we start with the dispersion relation of the non-interacting model,
\be
	E_\pm = \pm 4t\sqrt{\sin^2k_x + \sin^2k_y}.
\ee
We then replace each Majorana operator $\gamma^{\text{e}/\text{o}}$ with a combination of two slowly varying Majorana fields $\chi^{\text{e}/\text{o},\pm}$, according to
\be \label{eq1:1}
	\gamma^{\text{e}/\text{o}}(\vec{r}) \approx 2\sqrt{2}\Lambda_0^{-1}[\chi^{\text{e}/\text{o}+}(\vec{r})+ (-1)^x\chi^{\text{e}/\text{o}-}(\vec{r})].
\ee
These fields $\chi^\pm$ consist of the momenta modes of $\gamma$ near the two Dirac points of the non-interacting theory, which occur at $\vec{k}=(0,0)$ and $\vec{k}= (\pi/a,0)$. The coefficient $\Lambda_0^{-1} = a$ is the lattice spacing, and its inverse defines a bare energy cutoff of the theory. To derive the continuum limit, we Taylor expand the quadratic and quartic pieces of (\ref{1:1}). We expand the quartic operator to two derivatives, while keeping only leading order quadratic terms, since the underlying symmetry of the lattice model forbids any quadratic operator from breaking the U(1) symmetry (as proven in Appendix \ref{app:sym}). The resulting Hamiltonian density is
\be \label{eq1:2}
	\fH = 4it a \sum_\pm \Big[  \pm \chi^{\text{e}\pm}\partial_x\chi^{\text{e}\pm} \mp \chi^{\text{o}\pm}\partial_x\chi^{\text{o}\pm}
+ 2\chi^{\text{e}\pm}\partial_y\chi^{\text{o}\pm}\Big] + 64g\Lambda_0^{-4}\fH_{\text{int}}
\ee
where 
\be \label{eq1:99}
	 \mathcal{H}_{\text{int}} = 	
	- 4\Lambda_0^2\cem\cep\com\cop - \sum_{s,s' = \pm  }ss'\chi^{\text{e}s}\partial_x\chi^{\text{e}s}\chi^{\text{o}s'}\partial_x\chi^{\text{o}s'}
	+ 2\partial_y(\cem\cep)\partial_y(\com\cop)
\ee
\[
	+ 2\cem\cep
	\partial_x\com\partial_x\cop 
	+ 2
	\partial_x\cem\partial_x\cep\com\cop
	+ \partial_x(\cem\cep)\partial_x(\com\cop).
\]
We introduce two-component Majorana fermions $\chi^+ := (\cep,\cop)^{\text{T}}$ and  $\chi^- := (\com,\cem)^{\text{T}}$, so that the quadratic part of (\ref{eq1:2}) becomes
\be
	 4ita\sum_\pm \chi^{\pm T}[\sigma^z\partial_x + \sigma^x\partial_y]\chi^\pm.
\ee
These two-component Majorana fermions satisfy the canonical anti-commutation relations $\{\chi^i(\vec{r}),\chi^j(\vec{r'}) \} = \delta^{ij}\delta(\vec{r}-\vec{r'})$. The imaginary time Lagrangian density corresponding to (\ref{eq1:2}) is
\be \label{eq1:3}
	\fL = \sum_\pm \bar \chi^{ \pm}\gamma^\mu\partial_\mu \chi^\pm + 64g\Lambda_0^{-4}\fH_{\text{int}}.
\ee
We've set the velocity $v=4ta$ to unity, used the gamma matrices (\ref{eq:euclidgamma}), and defined $\bar\chi^\pm :=\chi^{\pm \text{T}} \gamma^0$.
  In order to identify any emergent U(1) invariance of (\ref{eq1:3}), we define a complex fermion $\psi$ according to 
\be \label{eq1:4}
	\psi = \chi^+ + i \chi^- = \begin{pmatrix} \cep + i\com \\
	\cop + i\cem \\
	\end{pmatrix}.
\ee

In this language, the most relevant U(1) breaking operator in (\ref{eq1:3}) is
\be \label{eq1:5}
	16g\Lambda_0^{-4}\Big( \psi_1\psi_2[\partial_x\psi_1\partial_x\psi_2 - \partial_y\psi_1\partial_y\psi_2] + \text{h.c.}\Big)
\ee
Including this term, the low energy field theory describing (\ref{1:1}) is
\be \label{eq1:10}
	\fL =  \bar\psi \gamma^\mu \partial_\mu \psi  + M\bar\psi\psi +   64g\Lambda_0^{-2}(\bar\psi\psi)^2
	+ 16g\Lambda_0^{-4}\Big(\psi_1\psi_2\partial_r\psi_1\partial_r\psi_2 + \text{h.c.}\Big)
\ee
where we've introduced the notation
\be
	\partial_r\psi_a\partial_r\psi_b := \partial_x\psi_a\partial_x\psi_b - \partial_y\psi_a\partial_y\psi_b.
\ee
and we've also introduced a fermion mass term: As shown in [\onlinecite{PhysRevB.96.125121}], when a second-neighbor hopping term is included, 
\be \label{1:4}
	H \to H + it_2\sum_{\substack{m,n \\
	s,s'= \pm1}}\gamma_{m,2n}\gamma_{m +s ,2n +s'},
\ee
time reversal symmetry is broken, and

\be
	\fL \to \fL + M\bar\psi\psi \hspace{20mm} M := 8t_2
\ee
 Since 
\be
	(\bar\psi\psi)^2 = -\psi_1^*\psi_2^*\psi_2\psi_1
\ee
we see from (\ref{eq1:10}) that $g>0$ corresponds to underlying physical interactions that are attractive. As a last comment, we note that the Nielson Ninomiya theorem\cite{Nielson1981} is not violated here, even though we have achieved a single Dirac fermion on the lattice, since the U(1) symmetry is only emergent, and not exact.

\subsection{Hubbard-Stratonovich Transformation}

In the absence of the U(1) breaking operator,  the interaction term in (\ref{eq1:10}) is proportional to $(\bar\psi\psi)^2$. In this case, we expect a massless boson to appear at the phase transitions $g_{\text{c},1}, g_{\text{c},2}$, whose expectation value provides the order parameter of the transition.\cite{ PhysRevB.96.125121, Klebanov2016} Such a boson can be introduced using a Hubbard-Stratonovich transformation. This procedure depends on the sign of the $(\bar\psi\psi)^2$ interaction: in the case of attractive interactions ($g>0$), a complex charge-2 boson is introduced, while in the case of repulsive interactions ($g<0)$, a real boson is introduced. To promote these bosonic variables to dynamical fields, we reduce the energy scale of the continuum theory from $\Lambda_0$ down to some  reduced scale $\Lambda \ll \Lambda_0$. Using the same symbols to denote these renormalized fields, we arrive at the following two imaginary time Lagrangian densities, depending on the sign of $g$:
\begin{itemize}
\item Repulsive Interactions ($g<0$):
\be \label{fl1}
	\fL_1 = \bar\psi \gamma^\mu\partial_\mu \psi  + (\partial_\mu\sigma)^2 + r^2\sigma^2 + \eta_1 \sigma\bar\psi\psi +\eta_2^2\sigma^4 
	+  h_1\left[\psi_1\psi_2\partial_r\psi_1\partial_r\psi_2 + \text{h.c.}\right]
\ee
\item Attractive Interactions ($g>0$):
\be  \label{fl2}
	\fL_2= \bar\psi\gamma^\mu\partial_\mu \psi +  M\bar\psi\psi+ |\partial_\mu\phi|^2 +m^2|\phi|^2 + \lambda_1 \left[\phi^*\psi^T C\psi + \text{h.c.}\right] + \lambda_2^2|\phi|^4 + \fL_2' 
\ee
\end{itemize}
where $C = i\gamma^0$ and 
\be
	\fL_2' :=h_2\psi_1\psi_2\partial_r\psi_1\partial_r\psi_2
	+ h_3\phi\partial_r\psi_1\partial_r\psi_2
	+h_4\phi[\partial_r^2\psi_1\psi_2 + \psi_1\partial_r^2\psi_2]
	+ \text{h.c.}
\ee 
We have only included a fermion mass in the case of attractive interactions; the phase transition for $g<0$ vanishes as soon as time reversal symmetry is broken, according to mean field theory.\cite{PhysRevB.96.125121} Note that in the case of attractive interactions, two additional U(1) breaking operators are generated during this renormalization procedure. Such terms do not occur for a real boson $\sigma$, since they violate an underlying $\frac{\pi}{2}$-rotation symmetry of the lattice, as shown in Appendix \ref{app:sym}. The Greek coupling constants $\{\lambda_i,\eta_i\}$ precede U(1) preserving operators, while the Latin coupling constants $\{h_i\}$ precede U(1) breaking operators. Equations (\ref{fl1}) and (\ref{fl2}) will be the starting point for all of our calculations that follow. We will assume that the symmetry breaking parameters $\{h_i\}$ and $M$ are small, so that the theories are close to their quantum critical points. This is not an unreasonable assumption for the lattice model: the U(1) breaking operators are superficially irrelevant, and are preceded by a factor of $\Lambda_0^{-4}$. At a reduced cutoff $\Lambda \ll \Lambda_0$, the coupling constants will be suppressed by four factors of $\Lambda/\Lambda_0$. Of course, this argument is incomplete, as it ignores higher order renormalization effects. If the $\{h_i\}$ and $M$ are not small, their flow will depend on the the presence of additional fixed points in parameter space.

We have assumed that under this renormalization, the velocities of the boson and fermion flow to a common value. This has been shown to be the case in the U(1) invariant versions of these models, and to linear order in $M$ and $\{h_i\}$, we expect the same result to hold.\cite{Lee2007, Ponte2014} The irrelevance of Lorentz breaking operators has also been established for fermion-boson models on the honeycomb lattice.\cite{PhysRevLett.97.146401, PhysRevB.79.085116, Herbut:2009vu} The fermion and boson velocities would be identical if Lorentz invariance was exact.

In [\onlinecite{PhysRevB.96.125121}], the nature of the transitions at $g_{\text{c},1}$ and $g_{\text{c},2}$ was predicted using the U(1) symmetric versions of (\ref{fl1}) and (\ref{fl2}), and invoking universality. In the fermion-boson models, the transitions are driven by reducing the squared boson mass, and letting it change sign. The U(1) symmetric version of (\ref{fl1}) was considered in [\onlinecite{Klebanov2016}], and the transition was shown to correspond to that of the Gross-Neveu model, with spontaneous breaking of the $\mathbb{Z}_2$ symmetry
\be
	\sigma \to -\sigma \hspace{5mm} \bar\psi\psi \to -\bar\psi\psi
\ee
It is not the Ising transition, because an additional massless fermion field $\psi$ is present. The U(1) version of (\ref{fl2}) involving the charge-2 boson $\phi$, (\ref{fl2}), has been studied as well\cite{Thomas2005, Lee2007,Ponte2014, Zerf2016, Klebanov2016} and the transition is known to exhibit $\fN=2$ supersymmetry when $M=0$. This should not be confused with the $\fN=1$ supersymmetry that is present in [\onlinecite{Grover2014}].

\subsection{Symmetry Constraints on U(1) Breaking Operators} \label{section:symmetry}

To complete this section, we comment on the symmetries of (\ref{eq1:10}). The authors of [\onlinecite{PhysRevB.96.125121}] identified various exact symmetries of the lattice Hamiltonian (\ref{1:1}), which must be obeyed at the continuum level. We label them $C$ for charge conjugation, $P$ for parity, and $R$ for $\frac{\pi}{2}$-spatial rotation. Explicitly, they are:
\begin{align} 
C: & \hspace{10mm} \psi(x,y) \mapsto \psi^*(x,y)  \label{eq1:6} \\
P:  & \hspace{10mm} \psi(x,y) \mapsto -i\gamma^1\psi^*(-x,y)  \label{eq1:8}\\
R: & \hspace{10mm} \psi(x,y) \mapsto e^{-\frac{i \pi }{4}} e^{\frac{i\pi}{4}\gamma^0} \psi(-y,x)
 \label{eq1:9}
\end{align}
Additionally, in the special case of $t_2= M =0$, the model is also invariant under time reversal, $T$:
\be
T:  \psi(x,y) \mapsto -\gamma^0 \psi^*(x,y),  \hspace{5mm} i \mapsto -i \label{eq1:7} 
\ee

In Appendix \ref{app:sym}, we show how these symmetries limit which U(1) breaking operators can be generated.

\section{Modified Epsilon Expansion} \label{section:modified}

In this paper, we seek to calculate the beta functions of various operators using an $\epsilon$-expansion. In both (\ref{fl1}) and (\ref{fl2}), the upper critical dimension of the U(1) invariant fermion-boson operator is four, and so we should consider these theories in $d=4-\epsilon$ dimensions, for $\epsilon \ll1$. However, this approach is met with difficulties, since $\bar\psi\psi$ is no longer a Lorentz scalar in four dimensions. Instead, it is a component of the 4-vector
\be
	A = \begin{pmatrix} \bar\psi \gamma^\mu \psi \\
	\bar\psi \psi \\
	\end{pmatrix}.
\ee
as shown in Appendix \ref{app:masslorentz}. The presence of $\bar\psi\psi$, either as a fermion mass term in (\ref{fl2}) or as a Yukawa coupling $\sigma\bar\psi\psi$ in (\ref{fl1}), will lead to the generation of additional Lorentz breaking operators through renormalization. In Appendix \ref{app:modified}, we explain this further, and write down the fermion and boson propagators in these non-Lorentz invariant theories.

Of course, this is not the first time an $\epsilon$-expansion has been attempted on these models. In the case of attractive  interactions, the conventional approach is to relate (\ref{fl2}) to the Nambu-Jona-Lasinio model in four dimensions, involving a 4-component Majorana fermion $\chi$, and two real bosons $\phi_1$ and $\phi_2$.\cite{Zerf2016, Klebanov2016, PhysRevD.96.096010} The interaction term in this model is
\be
	\bar\chi (\phi_1 + i\gamma_5 \phi_2)\chi
\ee
where $\gamma_5$ is the fifth gamma matrix in four dimensions. In the massless case, this theory possesses a continuous U(1) chiral symmetry:
\be
	\chi \to e^{i\alpha \gamma_5}\chi \hspace{10mm} \phi \to e^{-2i\alpha}\phi
\ee
In the Majorana representation, $\gamma_5$ is pure imaginary, so that this transformation leaves the Majorana real. In three dimensions, this model corresponds to the U(1) version of (\ref{fl2}), with the chiral U(1) mapping to the charge U(1) symmetry in the three dimensional theory. However, since a Majorana mass breaks the chiral U(1), we are unable to adopt this approach to our model when a fermion mass term is present.

Another popular approach in the literature, for the U(1) versions of both (\ref{fl1}) and (\ref{fl2}), is to extend the theory to one of $N$ Dirac fermions in four dimensions, and then continue $N \to \frac{1}{2}$ in the $\epsilon$-expansion.\cite{Klebanov2016} This approach is difficult to justify, since a four dimensional Dirac mass does not correspond to a three dimensional Dirac mass in this limit. See for instance, [\onlinecite{DiPietro:2017kcd}]. Instead, the four dimensional mass couples different chiral sectors together, as shown in Appendix \ref{app:masslorentz}. Using a change of basis, we can decouple the sectors, but in this case the three dimensional masses occur with opposite signs, and the limit $N \to \frac{1}{2}$ is ill-defined.

Therefore, we are forced to develop a new approach, which we call the `modified $\epsilon$-expansion', in order to calculate renormalization group functions in these theories. In the end, this approach will agree with the naive $N \to \frac{1}{2}$ limit in a conventional $\epsilon$-expansion, but is arguably more reliable, since it keeps the form of all operators fixed as $d$ is continued back to three dimensions. Perhaps there is a simple argument justifying the $N \to \frac{1}{2}$ limit, but we haven't been able to produce one. 

\subsection{An Expansion in $d=3+(1-\epsilon)$ Dimensions}

In Appendix \ref{app:modified}, we show that in four dimensions, the fermion and boson propagators receive Lorentz breaking corrections, due to the presence of $\bar\psi\psi$. These corrections lead to additional contributions in renormalization group functions that are unphysical, since the three dimensional theory is Lorentz invariant.  The modified $\epsilon$-expansion is a method to extract only the Lorentz invariant contributions in our Feynman diagram calculations. It isolates the Lorentz breaking direction (`$p_3$' in momentum space), and shrinks it to zero extent in the $\epsilon \to 1$ limit. To understand this, recall that the conventional $\epsilon$-expansion is carried out at the level of internal momentum integrals. In a Lorentz invariant theory, all momentum integrals will have the structure
\be
	\int\frac{d^4p}{(2\pi)^4} F(p) 
\ee
for some function $F$ depending only on the magnitude of momentum. Now, we continue from four to $d$ dimensions, by writing
\be
	\int \frac{d^4p}{(2\pi)^4}F(p) \to \int\frac{d^dp}{(2\pi)^d} F(p) = \Omega_d \int dpp^{d-1} F(p)
\ee
where $\Omega_d $ is the surface area of the sphere $S_{d-1}$. Both $\Omega_d$ and the radial integral are well-defined as functions of a continuous parameter $d$.

Now, let us turn to our non-Lorentz invariant theory, which has propagators modified by terms proportional to $\frac{p_3}{p^2}$. To lowest order in $p_3^2$, any Lorentz breaking contribution to a momentum integral will have the structure 
\be
	\int\frac{d^4p}{(2\pi)^4} F(p) p_3^2 
\ee
since odd powers of $p_3$ vanish by the symmetric integration. Higher powers of $p_3^2$ will be at least quadratic in the Lorentz breaking parameters, and can be dropped, as argued in Appendix \ref{app:modified}). In the conventional $\epsilon$-expansion, we would now promote $p$ to a $d$-dimensional vector, write $p_3^2 = p^2F'(\theta_i)$ in terms of spherical coordinates, and find some nonzero contribution. But this is unphysical, since all Lorentz breaking contributions should vanish when we return to the three dimensional theory. Instead, we promote $p$ to a $3+d'$ dimensional vector, and $p_3$ to a $d'$ dimensional vector, with $d'= 1-\epsilon$:
\be
	\int\frac{d^4p}{(2\pi)^4} f(p) p_3^2
	\to \int\frac{d^{3+d'}p}{(2\pi)^{3+d'}} F(p) |p_3|^2 = 
	\sum_{i=3}^{2+d'}  \int\frac{d^{3+d'}p}{(2\pi)^{3+d'}} F(p) p_{i}^2
\ee
\be
	= d' \int\frac{d^{3+d'}p}{(2\pi)^{3+d'}} F(p) p_{1}^2
	=\frac{d'}{3+d'}\Omega_{3+d'}
	 \int dp p^{3+d'+1}F(p)
\ee
In the limit $\epsilon\to1$, $d'\to0$, this integral vanishes. Since this applies to \emph{all} Lorentz breaking contributions to the Feynman diagrams, the modified $\epsilon$-expansion amounts to replacing the propagators in (\ref{eq:2}, \ref{eq:3}) with their Lorentz invariant pieces, and carrying out the conventional $\epsilon$-expansion.

If we instead promoted $\psi$ to a Dirac fermion in four dimensions, we would have Lorentz invariant propagators from the outset, with the $\gamma^\mu$ in (\ref{eq:2}) replaced with 4x4 gamma matrices. At the level of Feynman diagrams, our calculations would agree with those of the modified $\epsilon$-expansion, if we replaced $\tr(1_F)  \to 2$, which amounts to the naive $N \to \frac{1}{2}$ limit in the conventional $\epsilon$-expansion.

\section{Renormalization of U(1) Breaking Operators} \label{chapter:u1}

In this section, we determine the relevance of the U(1) breaking operators present in (\ref{fl1}) and (\ref{fl2}) to one loop order in the modified $\epsilon$-expansion, using Wilsonian renormalization. Reducing the energy cutoff from $\Lambda$ to $b^{-1}\Lambda$, for $b>1$, the coupling constants obey the following equations, in terms of renormalization constants $\{Z_i\}$:
\be \label{eq:oneloop:h1}
	h_1 = h_{1,0}Z_{h_1}Z_\psi^{-2} b^{-d}\hspace{10mm}
	h_2  =h_{2,0} Z_{h_2}Z_\psi^{-2}  b^{-d}
\ee
\be \label{eq:oneloop:h4}
	h_3  =h_{3,0} Z_{h_3}Z_\phi^{-1/2}b^{-d/2}
\hspace{10mm}
	h_4  =h_{4,0} Z_{h_4}Z_\phi^{-1/2} b^{-d/2}
\ee
Here $d$ is the spacetime dimension, which is 3 for the models of interest, and $4-\epsilon$ in the modified $\epsilon$-expansion.

\subsection{Attractive Interactions} 
In Appendix \ref{app:u1:attractive}, we calculate these renormalization constants for the case of attractive interactions, (\ref{fl2}). Our results are given in (\ref{eq:z1}-\ref{eq:z2}). Differentiating these expressions with respect to $\log b$, we find
\begin{align}
\beta_M 
 = &  M\left[  1 -  \epsilon\right] \\ 
\beta_{h_2} 
= & -h_2\left[ 4-\frac{\epsilon}{3}\right]
\\ 
\beta_{h_3} 
= &  -h_3\left[ 2 + \frac{6h_2\sqrt{\epsilon} }{\sqrt{3}(4\pi)h_3} \right]
  \\ 
\beta_{h_4} 
= & -h_4\left[ 2 - \frac{2 h_2 \sqrt{\epsilon}}{h_4 4\pi \sqrt{3}} - \frac{2 \epsilon(2h_4 - h_3)}{3h_4}\right]
 \\
\end{align}
 Above, we are using the convention
 \be
 	\beta_x := \frac{dx}{d\log b}.
\ee

Since $\beta_{h_2}$ is only a function of $h_2$, and is negative for $\epsilon =1$, we conclude that $h_2$ flows to zero at large length scales, independent of $h_3$ and $h_4$. This implies that $\beta_{h_3}$ is also negative at large length scales, so that $h_3 \to 0$. Finally, we are left with
\be
	\beta_{h_4} \to -h_4\left[ 2 - \frac{4\epsilon}{3}\right] \to -\frac{2}{3}h_4 <0
\ee 
so that $h_4$ also flows to zero. Therefore, at the critical point $g_{\text{c},2}$, all U(1) breaking operators are irrelevant.  Meanwhile, the fermion mass operator is marginal at one loop, and requires a higher order calculation. In the next section, we set the U(1) breaking operators to zero, and use supersymmetry to determine the relevance of a fermion mass to four loop order.

\subsection{Repulsive Interactions}
In Appendix \ref{app:u1:repulsive}, we calculate the renormalization constants for the case of repulsive interactions, (\ref{fl1}). Using (\ref{eq:z3}), we find 
\be
	\beta_{h_1} = -h_1\left[ 4 - \frac{3}{4}\epsilon\right]
\ee 
at the phase transition $g_{\text{c},1}$. Therefore, to one loop order, the U(1) breaking operator is irrelevant, and the phase transition falls into the Gross-Neveu universality class, as predicted in [\onlinecite{PhysRevB.96.125121}].

\section{Relevance of the Fermion Mass Operator} \label{chapter:mass}

In this section, we determine the relevance of the fermion mass operator in (\ref{fl2}) beyond one loop order. We treat $M$ as a small parameter, so that terms $\fO(M^2)$ will be dropped. We also neglect all U(1) breaking operators, since these were shown to be irrelevant in the previous section.

A straightforward, but tedious approach to the problem is to calculate all two loop diagrams in the modified $\epsilon$-expansion. The results of this calculation can be found to $\fO(\epsilon^2)$ in Appendix \ref{app:twoloop}. A more efficient approach is to relate the fermion mass beta function to the \emph{stability critical exponent} in the massless theory, which allows us to go to $\fO(\epsilon^4)$, using the following identity:
\be \label{3:1}
	\beta_M = M\left[1 - \omega\right] \hspace{10mm} \omega:= \frac{d}{d\lambda_1^2 }\frac{ d \lambda_1^2}{d\log \mu}\Big|_{\lambda_1= \lambda_1^*, \text{(massless)}}
\ee 
In words, $\omega$ is the derivative of the beta function for $\lambda_1^2$ in the massless theory, evaluated at the critical point. The proof of equation (\ref{3:1}) closely follows the derivation of the identity 
\be \label{eq:bosonmass}
	\beta_{m^2} = m[2 - \omega]
\ee
for the boson mass operator $m^2|\phi|^2$ in [\onlinecite{Zerf:2016fti}], and relies on the underlying supersymmetry of the massless theory. The relation (\ref{eq:bosonmass}) was originally given in [\onlinecite{Thomas2005}]. We now prove this identity using the superspace formalism. In Appendix \ref{app:susy}, we review this formalism, and show that at the critical point $g_{\text{c},2}$, the real time version of (\ref{fl2}) in $2+1$ dimensions can be written in terms of a chiral superfield, $\Phi$:
\be
	\int d^2\theta d^2\bar\theta \Phi^\dag \Phi +\frac{\lambda}{3}\left( \int d^2\theta \Phi^3 + \int d^2\bar\theta{\Phi^\dag}^3\right)
	= \partial_\mu \phi^* \partial_\nu\eta^{\mu\nu} \phi + i \bar\psi \gamma^{\mu}_R\partial_\mu \psi
	- \lambda^2|\phi|^4 - \lambda\left(\phi\psi^T C \psi + \text{h.c.}\right)
\ee
where $\gamma^\mu_R = \{-\gamma^0, i\gamma^1, i\gamma^2\}$ are real space gamma matrices, and $\eta^{\mu\nu} = \text{diag}(1,-1,-1)$. We have dropped all U(1) breaking operators, since they have been shown to be irrelevant in Section \ref{chapter:u1}. A fermion mass can be introduced by adding the following expression to (\ref{eq:susy:result}):
\be
	-\int d^2\theta d^2\bar\theta 2M \Phi^\dag \theta \bar\theta \Phi
	= -4M\int d^2\theta d^2\bar\theta \bar\theta \bar\psi \theta\bar\theta\theta\psi = -M\bar\psi\psi
\ee 

To linear order in $M$, this addition can be compensated by rescaling the superfield,
\be
	\Phi \to (1 + M \theta\bar\theta)\Phi
\ee
which shifts the coupling $\lambda$ accordingly:
\be \label{apeq:susy:4}
	\lambda \to \tilde\lambda(M):= \lambda + 3M \theta\bar\theta
\ee

In other words, the massive theory with coupling $\lambda$ is equivalent to the massless theory with coupling $\tilde\lambda$. Now, to access the scaling dimension of $\bar\psi\psi$, we require the notion of bare and renormalized fields and masses. We write the bare theory in terms of bare $\Phi_0$ and bare $M_0,\lambda_0$:
\be \label{apeq:susy:5}
	\fL_{\text{bare}} = \int d^2\theta d^2\bar\theta \Phi_0^\dag (1 -2M_0 \theta \bar\theta)\Phi_0  + \int d^2\theta \frac{\lambda_0}{3}\Phi_0^3 + \int d^2\bar\theta \frac{\lambda_0}{3}[\Phi_0^\dag]^3
\ee 
and the renormalized theory in terms of $\Phi$ and $M\mu, \lambda \mu^{\epsilon/2}$:
\be \label{apeq:susy:3}
	\fL = \int d^2\theta d^2\bar\theta \tilde Z \Phi^\dag (1 -2M \mu \theta \bar\theta)\Phi + \int d^2\theta \frac{\lambda \mu^{\epsilon/2}}{3}\Phi_0^3 + \int d^2\bar\theta \frac{\lambda \mu^{\epsilon/2}}{3}[\Phi_0^\dag]^3\ee 
Here the renormalization scale $\mu$ has been introduced so that $M$ and $\lambda$ are dimensionless. Notice that there is no renormalization constant $Z_\lambda$ -- this follows from SUSY nonrenormalization theorems.\cite{Seiberg:1993vc, Grisaru:1979wc} In the massless theory, we can write down an equation similar to (\ref{apeq:susy:3}), replacing $\tilde Z$ with some other renormalization constant $Z$. In general, these two functions will be different; however, using (\ref{apeq:susy:4}), we have
\be
	\tilde Z(\lambda) = Z(\tilde \lambda) = Z(\lambda)\left[ 1 + 3M \mu \theta \bar\theta \lambda \frac{\partial \log Z}{\partial\lambda}\right]  + \fO(M^2)
\ee
Using this and comparing (\ref{apeq:susy:5}) to (\ref{apeq:susy:3}), and we find the relation
\be \label{apeq:susy:6}
	M = M_0 \mu^{-1}\left[ 1 - \frac{3}{2}\lambda \frac{\partial \log Z}{\partial\lambda}\right]^{-1}
\ee
Differentiating with respect to $\log \mu$, we find
\be \label{apeq:susy:7}
	-\beta_M := \frac{\partial M}{\partial \log \mu} = M\left[ -1 - 3\lambda^2 \frac{\partial \gamma}{\partial\lambda^2}\right]
\ee
where $\gamma = -\frac{\partial Z}{\partial \log \mu}$ is the anomalous dimension of the fermion in the massless theory. The unconventional negative sign is introduced so that these functions agree with their Wilson counterparts. Now, in the supersymmetric theory, $\gamma$ can be rewritten in terms of the beta function of $\lambda^2$, since
\be
	\lambda_0^2 = \lambda^2 \mu^{-\epsilon}Z(\lambda)^3
\ee
because the superpotential is not renormalized. The beta function is
\be
	-\beta_{\lambda^2} = -\frac{d\lambda^2}{d\log \mu} = \lambda^2[-\epsilon  - 3\gamma]
\ee
Differentiating with respect to $\lambda^2$, and using the fact that to $\fO(\epsilon^4)$,  the value of $\gamma$ at the SUSY point ($ =:\lambda_*$) is\cite{PhysRevD.96.096010}
\be
	\gamma(\lambda_*) = -\frac{\epsilon}{3}
\ee
 we have
\be
	-\frac{d\beta_{\lambda^2}}{d\lambda^2} = -\epsilon - 3\gamma(\lambda_*)  -3\lambda_*^2 \frac{\partial \gamma}{\partial\lambda^2} = -3\lambda_*^2\frac{\partial \gamma}{\partial \lambda^2}
\ee

Comparing this to (\ref{apeq:susy:7}), we find 
\be
	 \beta_M = M\left[ 1  - \frac{d\beta_{\lambda^2}}{d\lambda^2}\right]
\ee
proving (\ref{3:1}).  In [\onlinecite{PhysRevD.96.096010}], $\omega$ has been evaluated in the massless theory to four loop order:
\be \label{eq:omega}
	\omega = \epsilon - \frac{\epsilon^2}{3} + \left( \frac{1}{18} + \frac{2\zeta_3}{3}\right)\epsilon^3 
	+ \frac{1}{540}\left( 420\zeta_3 + 1200\zeta_5 - 3\pi^4 + 35\right)\epsilon^4 +\fO(\epsilon^5)
\ee
Using Pad\'e extrapolation,\cite{Klebanov2016} the authors of [\onlinecite{PhysRevD.96.096010}] found the values $\omega = 0.872$ and $\omega=0.870$, depending on which Pad\'e approximant is used. In [\onlinecite{PhysRevLett.115.051601}], the value $\omega = .910$ was obtained using the conformal bootstrap. In all three approaches, 
\be 
	\beta_M = M\left[ 1 - \omega\right]
\ee 
is positive, and the fermion mass operator is relevant. Therefore, at the phase transition $g_{\text{c},2}$, a small time reversal breaking perturbation will destroy the emergent supersymmetry. The resulting universality class is determined in the following subsection. In passing, we note that our explicit two loop results, calculated using dimensional regularization, agree with (\ref{3:1}) and (\ref{eq:omega}) to $\fO(\epsilon^2)$ (see Appendix \ref{app:twoloop}).

\subsection{The Effect of a Relevant Fermion Mass Operator}

Since the fermion mass operator is relevant, a large mass will be generated near the critical point. At energy scales $\ll M$, the fermion degrees of freedom can be integrated out completely. In Appendix \ref{app:largem}, we show that in this case, the low energy theory near the critical point $g_{\text{c},2}$ has the following structure
\be \label{eq3:12}
	\fL = |\partial_\mu \phi|^2 + m^2|\phi|^2 +  \rho |\phi|^4 + \tilde \rho (\phi_4 + \phi_4^*) \hspace{10mm}  \tilde \rho \ll \rho
\ee

This model was studied in [\onlinecite{2000PhRvB..61.3430O}] and  [\onlinecite{2000PhRvB..6115136M}] using $\epsilon$-expansion techniques and in [\onlinecite{PhysRevLett.99.207203}] using Monte Carlo, 
where it was shown that $\tilde \rho$, which lowers the symmetry from U(1) to $Z_4$, is irrelevant in 3 spacetime dimensions and the critical point is the XY one. Therefore, once a fermion mass is present, the universality class of $g_{\text{c},2}$ will change from $\fN=2$ SUSY to the conventional XY transition. 

\section{Conclusion} \label{section:conclusion}

In this work, we have shown that the emergent U(1) symmetry present at the critical points of the Majorana-Hubbard model is preserved when U(1) breaking corrections are taken into account. Moreover, we have shown that a fermion mass term, generated by a time reversal breaking perturbation, is a relevant operator at four loops in the $\epsilon$-expansion. These results suggest that in the case of repulsive interactions, the Majorana-Hubbard model has a critical point in the Gross-Neveu universality class, and in the case of attractive interactions, the model has a critical point in the $\fN=2$ supersymmetric universality class for a time reversal invariant system. When time reversal symmetry is broken, we have shown that the phase transition instead falls in the conventional XY universality class. These results agree with the classification of Affleck et. al.\cite{PhysRevB.96.125121}. Numerical confirmation of these predictions remains a major open challenge.

\begin{acknowledgments}
We would like to thank Igor Herbut, Igor Klebanov, Joseph Maciejko, Dmitry Pikulin, Armin Rahmani and Hennadii Yerzhakov for helpful discussions. This research was supported by NSERC of Canada and the QuEST Program of the Stewart Blusson Quantum Matter Institute.
\end{acknowledgments}

\bibliography{MSC_thesis}
\bibliographystyle{unsrt}

\appendix

\section{Symmetry Constraints on U(1) Breaking Operators} \label{app:sym}

In this appendix, we show how the symmetries (\ref{eq1:6}-\ref{eq1:9}) restrict the form of various U(1) breaking operators.

\subsection{Quadratic Operators} \label{quadratic}

The most general U(1) breaking quadratic operator (with or without derivatives) is of the form
\be
	\psi^T A \psi + \psi^\dag A^\dag \psi^*
\ee 
 for some differential operator $A(x,y)$. Under $C$, 
\be \label{c0}
	C: \psi^T A \psi + \psi^\dag A^\dag \psi^*
	\mapsto \psi^T A^\dag \psi + \psi^\dag A \psi^*
\ee
which forces $A$ to be Hermitian. Under $R$, 
\be \label{c1}
	R: \psi^T A(x,y) \psi \mapsto
	-\frac{i}{2} \psi^T(1-i\sigma_y) A(-y,x)(1+i\sigma_y)\psi.
\ee
The right hand side of (\ref{c1}) cannot appear for nonzero $A$, since it is anti-Hermitian, and violates (\ref{c0}). Therefore, no charge 2 operator is allowed by symmetry.

\subsection{Quartic Operators} \label{quartic1}

\subsubsection{One-Derivative Quartic Operators} \label{quartic2}

A four-Fermi operator involving a single derivative can only have charge 0 or $\pm2$: terms with charge $\pm 4$ include at least three fermi fields without derivatives, and vanish by Fermi statistics. Since $R$ is a combination of spatial rotation by $\frac{\pi}{2}$ and U(1) rotation by $-\frac{\pi}{4}$, these two possibilities require, respectively, a derivative operator that transforms trivially or one that transforms with a prefactor of $i$. Of these, only the latter exists:
\be
	\partial_x+i\partial_y
\ee
but such an operator breaks $CP$.

\subsubsection{Two-Derivative Quartic Operators}

Repeating the previous argument, the derivative operator of a charge 2 four-fermi term must transform with a factor of $i$ to satisfy $R$ symmetry. This is not possible for a generic two-derivative operator $A_{ab}\partial_a\partial_b$, ruling out charge 2 operators. Charge 4 terms require a derivative operator that transforms with a prefactor of $-1$ to be invariant under $R$. By Fermi statistics, the two derivatives must act on separate Fermi fields, so the most general  operators are 
\be
	\psi_1\psi_2[\partial_x\psi_1\partial_x\psi_2 - \partial_y\psi_1\partial_y\psi_2]
\ee
and
\be
	\psi_1\psi_2[\partial_x\psi_1\partial_y\psi_2 - \partial_y\psi_1\partial_x\psi_2]
\ee
Of these, only the former is allowed, since the latter breaks $CP$. Therefore, the U(1) breaking operator appearing in (\ref{eq1:3}) is the only possible term with two or less derivatives.

\subsection{Fermion-Boson Operators}

In the case of attractive interactions, a complex boson $\phi \sim \psi_1\psi_2$ is introduced. Using (\ref{eq1:9}), we see that
\be
	R: \phi(x,y) \to i\phi(-y,x)
\ee
Since $\psi^T C\psi$ also picks up a factor of $i$ under $R$, the following two derivative, U(1) breaking operators are invariant under $R$-symmetry:
\be
	\phi[\partial_x\psi_1\partial_x\psi_2 -\partial_y\psi_1\partial_y\psi_2] + \text{h.c.}
\ee
and
\be
	\phi\left[ (\partial_x^2-\partial_y^2)\psi_1\psi_2 + \psi_1(\partial_x^2-\partial_y^2)\psi_2\right] + \text{h.c.}
\ee
It is easy to check that the remaining symmetries (\ref{eq1:6} - \ref{eq1:9}) also leave these operators invariant.

In the case of repulsive interactions, a real boson $\sigma \sim \bar\psi\psi$ is introduced, which is invariant under $R$:
\be
	R: \sigma(x,y) \to \sigma(-y,x)
\ee
Using the above constraints on pure fermion operators, the most relevant U(1) breaking fermion-boson operator is then 

\be
	\sigma^2 \psi_1\psi_2[\partial_x\psi_1\partial_x\psi_2 - \partial_y\psi_1\partial_y\psi_2]
\ee
which is too irrelevant for our considerations.

\section{Weyl Fermions in Four Dimensions} \label{app:masslorentz}

In four dimensions, the Dirac Lagrangian is 
\be \label{eq:weyllagrangian}
	\fL = i\bar\Psi \Gamma^a \partial_a \Psi \hspace{10mm} \Psi = (\psi_R \,\,\, \psi_L)^T \hspace{10mm} a = 0,1,2,3 
\ee
The gamma matrices are in the Weyl basis, and can be written in terms of two sets of Pauli matrices $\{\sigma_i\}$ and $\{\tau_i\}$:
\be
	\Gamma^0 = \tau_x\otimes \sigma_0 \hspace{10mm} \Gamma^k = i\tau_y \otimes \sigma_k 
\ee
where $\sigma_0 := 1$. These matrices satisfy
\be
	\{\Gamma^a, \Gamma^b \} = 2\text{diag}(1,-1,-1,-1).
\ee
Expanding (\ref{eq:weyllagrangian}), the $\psi_L$ sector can be written as
\be
	\fL_W = i\bar\psi\sigma_y\partial_0 \psi + i\bar\psi[\sigma_y\sigma_x \partial_1 + \partial_2 +\sigma_y\sigma_z\partial_3]\psi \hspace{10mm} \bar\psi := \psi^\dag\gamma^0 = \psi^\dag\sigma_y
\ee
where we've suppressed the `L' subscript, and inserted $\sigma_y^2=1$ in each term. By relabelling coordinates $\partial_2\leftrightarrow \partial_3$, and performing a Wick rotation, the imaginary time Lagrangian density for the Weyl fermion is
\be \label{eq:weyl2}
	\fL = \bar\psi[\partial_\mu\gamma^\mu + i\partial_3]\psi \hspace{10mm} \mu = 0,1,2
\ee
Since (\ref{eq:weyl2}) is a Lorentz scalar, and $(\partial_\mu,\partial_3)$ is a 4-vector, we see that $\bar\psi\psi$ is no longer invariant under the Lorentz group. This can also be seen explicitly, using the general form of a Lorentz transformation in the Weyl basis:\cite{Peskin:1995ev}
\be \label{eq:lortran}
	\Lambda(\alpha) = e^{\vec{\alpha}\cdot\vec{\sigma}} \hspace{10mm} \vec{\alpha} \in \mathbb{C}
\ee
Under this transformation, with $\gamma^0=\sigma_y$,
\be
	\bar\psi\psi \mapsto \psi^\dag e^{\vec{\alpha}^*\cdot\vec{\sigma}}\gamma^0 e^{\vec{\alpha}\cdot\vec{\sigma}}\psi
	= \bar\psi e^{-\vec{\alpha}^*\cdot\vec{\sigma}^T} e^{\vec{\alpha}\cdot \vec{\sigma}}\psi 
\ee

which does not equal $\bar\psi\psi$ for general $\vec{\alpha}$. It is only invariant under a subset of operators,
\[
	\{ e^{\lambda \sigma_x}, e^{\lambda \sigma_z}, e^{i\lambda \sigma_y} \}, \hspace{10mm} \lambda \in\mathbb{R}
\]
which generate the three dimensional Lorentz group.

\subsection{The Limit $N \to \frac{1}{2}$}
One idea to resolve the issue of breaking Lorentz invariance in the $\epsilon$-expansion is to promote $\psi$ to a Dirac fermion in four dimensions. If this Dirac theory can be decoupled into two Weyl sectors, then we may obtain the Weyl renormalization group functions by continuing $N$, the number of Dirac fermions, from 1 to $\frac{1}{2}$ in this theory. We now show that this limit is ill-defined. 

To generate the interaction term $\phi^*\psi^T C\psi$ in each Weyl sector, we consider following operator
\be
	i \phi^*\Psi^T \begin{pmatrix} C & 0 \\
	0 & -C \\
	\end{pmatrix} \Psi + \text{h.c.} \hspace{10mm} C = i\gamma^0
\ee
To show that it is Lorentz invariant, it is sufficient to consider $\psi^T C\psi$, since Lorentz transformations do not couple Weyl sectors in the Weyl basis. Using (\ref{eq:lortran}), 
\be
	\psi^TC\psi \mapsto \psi^T e^{\vec{\alpha}\cdot \vec{\sigma}^T} C e^{\vec{\alpha}\cdot\vec{\sigma}}
	= \psi^T C e^{-\vec{\alpha}\cdot\vec{\sigma}} e^{\vec{\alpha} \cdot\vec{\sigma}}\psi = \psi^TC\psi
\ee
under a general Lorentz transformation. Adding this interaction to the free Dirac Lagrangian density, we have
\be \label{eq:dirac2}
	\fL = \bar\Psi[ \partial_a\Gamma^a + M]\Psi + [i \phi^*\Psi^T \begin{pmatrix} C & 0 \\
	0 & -C \\
	\end{pmatrix} \Psi + \text{h.c.}]
\ee  
By rotating $\psi_R \to \gamma_0\psi_R$, so that both Weyl fermions propagate in the same direction, (\ref{eq:dirac2}) becomes 
\be
	\sum_{i=1}^2 \left[\bar\psi_i[\partial_\mu\gamma^\mu +i\partial_3]\psi_i + [i\phi^*\psi_i^TC\psi_i + \text{h.c.}]\right]
	+ M\bar\psi_L\psi_R + \bar\psi_R\psi_L
\ee
where we used (\ref{eq:weyl2}). The two Weyl sectors can be decoupled by introducing
\be
	\psi_\pm := \frac{1}{\sqrt{2}} (\psi_L \pm \psi_R).
\ee
This doesn't affect the interaction term, but it modifies the mass terms to
\be
	M[\bar\psi_L\psi_L - \bar\psi_R\psi_R]
\ee

This relative sign in the mass terms cannot be removed, implying that the two Weyl sectors are distinct. Any continuation of the Dirac number $N \to \frac{1}{2}$ would have to choose between one of these two distinct sectors, rendering the limit ill-defined.

\section{Lorentz Breaking Operators in Four Dimensions} \label{app:modified}
 In four dimensions, a two-component complex fermion is a Weyl fermion, with imaginary time Lagrangian density
\be \label{eq:flw}
	\fL_W = \bar\psi [\partial_\mu\gamma^\mu + i\partial_3]\psi \hspace{10mm} \mu = 0,1,2
\ee
This quadratic form can be derived from the Dirac Lagrangian in 4 dimensions, using $4\times 4$ Gamma matrices in the Weyl basis.  Since (\ref{eq:flw}) is a Lorentz scalar, and $(\partial_\mu, \partial_3)$ is a 4-vector, we see that the object $\bar\psi\psi$ is no longer invariant under the Lorentz group. Instead, it is a component of the 4-vector, 
\be
	A = \begin{pmatrix} \bar\psi \gamma^\mu \psi \\
	\bar\psi\psi \\
	\end{pmatrix},
\ee
that is contracted with $(\partial_\mu,\partial_3)$ in (\ref{eq:flw}). This creates difficulties when studying the fermion mass operator $M\bar\psi\psi$, as well as the Gross-Neveu interaction $\bar\psi\psi \sigma$ in (\ref{fl1}). While these operators are invariant under the three dimensional  Euclidean Lorentz group SO(3), they transform nontrivially under the full SO(4) Euclidean Lorentz group. As a consequence, additional operators that are invariant only under the SO(3) $\subset$ SO(4) subgroup can be generated, including (for $k \in\mathbb{Z_+}$)
\be
	\bar\psi (i\partial_3)^k \psi \hspace{10mm} |\partial^{k}_3\phi|^{2}\hspace{10mm} (\partial_3^k\sigma)^{2} \hspace{10mm}  (\phi\partial_3^k \phi^* + \text{h.c.}) \hspace{10mm}\sigma\partial_3^k\sigma 
\ee

We will only discuss the role of the most relevant operators, with $k=1$. Then in four dimensions, we should replace the Lagrangian densities (\ref{fl1}) and (\ref{fl2}) with the following:
\be \label{fl1:new}
	\fL_1' = \bar\psi[ \sld + i\partial_3 + if_1\partial_3]\psi + M\bar\psi\psi + (\partial_a \sigma)^2 + f_2(\partial_3\sigma)^2 +\eta_1\sigma \bar\psi \psi + \eta_2 \sigma^4
\ee
\[
	+ f_3\sigma\partial_3\sigma + \cdots 
\]
\be \label{fl2:new}
	\fL_2'= \bar\psi[ \sld + i\partial_3 + if_1\partial_3]\psi + M\bar\psi\psi + |\partial_a \phi|^2 + f_2|\partial_3\phi|^2 +\lambda_1 [ \phi \psi^T C \psi + \text{h.c.}]
\ee
\[
	 + \lambda_2^2|\phi|^2
	+ f_3 (\phi \partial_3 \phi^* + \text{h.c.})  + \cdots
\]

The `$\cdots$' represent the U(1) breaking operators present in (\ref{fl1}) and (\ref{fl2}), which are unchanged. Since the parameters $\{f_i\}$ are not present in the three dimensional model, they only appear in the four dimensional model after at least one renormalization step, and are suppressed by at least one factor of $M$ or $\eta_1$ (the Lorentz breaking operators in (\ref{fl1:new}) and  (\ref{fl2:new})). In either case, terms $\fO(f_i^2)$ and $\fO(f_i M)$ are beyond our order of approximation, and should be dropped from the calculations that follow.  

\subsection{Propagators} 

Inverting the quadratic forms in (\ref{fl1:new} \ref{fl2:new}), we find the following propagators, to linear order in $M$ and $f_i$:

\be
	G(p) = \langle \psi(p)\bar\psi(p)\rangle 
	= \frac{i\stp + f_1p_3 + M}{p^2}
-	2p_3(M+f_1p_3) \frac{i\stp + p_3 }{p^4}
\ee
where we've introduced a four dimensional  `slash notation'
\be
	\text{\st{A}} :=A_\mu\gamma^\mu - iA_3
\ee
We write this propagator as a sum of a Lorentz invariant $(G_1)$ and a non-Lorentz invariant $(G_2)$ part:
\be \label{eq:2}
	G(p)  = G_1(p) + G_2(p)
\ee
\be
	G_1(p)
	= \frac{ i\stp + M}{p^2}
\hspace{10mm}
	G_2(p) = \frac{p_3}{p^2}\left[ f_1 -2(M+f_1p_3) \frac{i\stp + p_3}{p^2}\right]
\ee
Only the first term is a Lorentz invariant.  Likewise, the  boson propagators are
\be \label{eq:3}
	D(p) = \langle \sigma(p)\sigma(-p)\rangle = \langle \phi(p) \phi^*(p)\rangle = D_1 + D_2 
\ee
where
\be
	D_1(p) = \frac{1}{p^2} \hspace{10mm} D_2(p) = - \frac{f_2 p_3^2}{p^4}  - \frac{f_3ip_3}{p^4}
\ee

\section{Renormalization of U(1) Breaking Operators} \label{app:u1}

In this appendix, we determine the relevance of the U(1) breaking operators present in (\ref{fl1}) and (\ref{fl2}) to one loop order in the modified $\epsilon$-expansion, using Wilsonian renormalization. To begin, we decompose fields into slow and fast components. Throughout our calculations, we consider all one loop diagrams that are $\fO(h_i), \fO(\lambda_i^2), \fO(\eta_i^2)$ and $\fO(M)$. We define the operator $*$ on momenta vectors $a,b$ as
\be
	a*b := a_xb_x - a_yb_y
\ee
and we use faint/bold propagator lines to denote slow/fast fields in our Feynman diagrams. We also use the notation  $\stp := \slp -ip_3$, introduced in Appendix \ref{app:modified}. From the outset, we set the boson masses to zero, since this marks the phase transitions of interest. All Feynman diagrams have been drawn using the package [\onlinecite{Ellis:2016jkw}].

\subsection{U(1) Breaking Operators with Attractive Interactions} \label{app:u1:attractive}

Using the modified $\epsilon$-expansion, the fermion and boson propagators are
\be
	G(p) = \frac{i\stp +M}{p^2} \hspace{10mm} D(p) = \frac{1}{p^2}
\ee
We use solid lines to represent the fermion propagators, and dashed lines to represent the boson propagators. An arrow is used to indicate the direction of charge; this charge is +1 for the fermion, and +2 for the boson. Finally, we include the operators of the external legs in the definitions of our Feynman diagrams. 

\subsubsection{Fermion Propagator}  

\begin{figure}[ht]
\begin{center}
\includegraphics[width=0.3\textwidth]{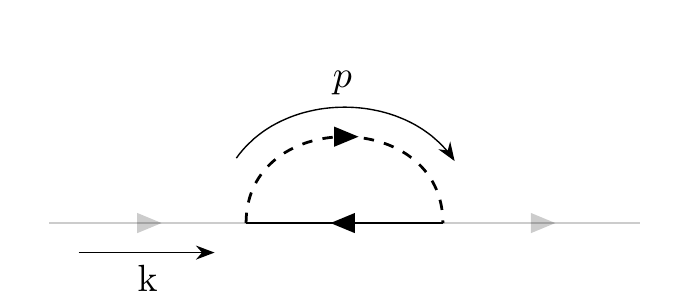}
\caption{Fermion self energy diagram in Wilson RG for $g>0$}
\label{oneloop:diag1}
\end{center}
\end{figure}

The single one loop diagram that renormalizes the fermion propagator to $\fO(h_i)$ is shown in Figure \ref{oneloop:diag1}. Including the external legs, it equals
\be
	= \int \frac{d^dk}{(2\pi)^d} \bar\psi_s(k) \Sigma_\psi(k)\psi_s(k)
\ee
where
\be
	\Sigma_\psi(k) =  -4\lambda_1^2 \int_f \frac{d^dp}{(2\pi)^d} D(p) C^T G^T(p-k)C
\ee 
and the $p$ integration is over the Wilson shell. Expanding to linear order in the slow momentum $k$, and replacing 
\be
	p\cdot k i\stp^\dag \to \frac{p^2}{d} i\stk^\dag,
\ee 
we find
\be
	\Sigma_\psi(k) = -i\stk^\dag 4\lambda_1^2\left[ 1 - \frac{2}{d}\right] \int_f \frac{d^dp}{(2\pi)^d} \frac{1}{p^4}
	- 4\lambda_1^2M \int_f \frac{d^dp}{(2\pi)^d} \frac{1}{p^4}
\ee 
Using
\be
	\int_f \frac{d^dp}{(2\pi)^d} \frac{1}{p^4}  = \Omega_d \int_{b^{-1}\Lambda}^\Lambda dp p^{d-5}
	 = \frac{2}{(4\pi)^{d/2} \Gamma(d/2)} \Lambda^{d-4}\delta l + \fO(\delta l^2)
\ee
for $b = e^{\delta l}$, we find the following renormalization constants for the fermion kinetic term and fermion mass term:
\be
	Z_\psi = 1 +  \frac{8\lambda_1^2}{(4\pi)^{d/2}\Gamma(d/2)} \left[ 1 - \frac{2}{d}\right] \Lambda^{-\epsilon}\delta l \hspace{10mm}
	Z_M = 1 - \frac{8\lambda_1^2}{(4\pi)^{d/2}\Gamma(d/2)}\Lambda^{-\epsilon}\delta l 
\ee

\subsubsection{Boson Propagator}

\begin{figure}[ht]
\begin{center}
\includegraphics[width=0.3\textwidth]{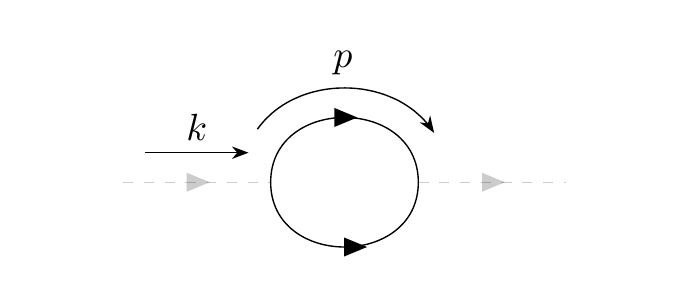}
\caption{Boson self energy diagram in Wilson RG}
\label{oneloop:diag2}
\end{center}
\end{figure}

The unique one loop diagram that renormalizes the boson propagator to linear order in $\fO(h_i)$ is shown in Figure \ref{oneloop:diag2}.
It equals
\be
	= \int \frac{d^dk}{(2\pi)^d}\phi_s^*(k)\Sigma_\phi(k)\phi_s(k)
\ee
where
\be
	\Sigma_\phi(k) = 2\lambda_1^2 \int_f \frac{d^dp}{(2\pi)^d} \text{tr}[ CG(p) CG^T(k-p)]
\ee
Since the phase transition occurs when the boson mass is tuned to zero, we isolate the terms proportional to $k^2$, to extract $Z_\phi$. We need not be concerned with the generation of terms proportional to $k_4$ only, since these drop out of the modified $\epsilon$-expansion. We find 
\be
	Z_\phi = 1 +\frac{8}{(4\pi)^{d/2}\Gamma(d/2)}\left[ 1- \frac{2}{d}\right] \lambda_1^2 \Lambda^{-\epsilon}\delta l.
\ee

\subsubsection{Renormalization of $h_2,h_3$ and $h_4$}
\begin{figure}[ht]
\begin{center}
\includegraphics[width=0.3\textwidth]{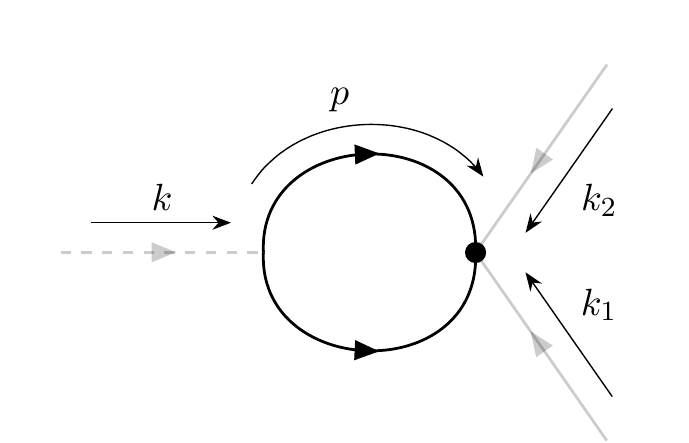}
\caption{First diagram renormalizing $h_3$ and $h_4$ in Wilson RG.}
\label{oneloop:fourloop}
\end{center}
\end{figure}

At one loop, there is no diagram renormalizing $h_2$, so
\be
	Z_{h_2} = 1
\ee
There are two diagrams that contribution to the renormalization of $h_3$ and $h_4$ at one loop. The first is shown in Figure \ref{oneloop:fourloop}, and equals
\be
= \int \frac{d^dk_1}{(2\pi)^d} \frac{d^2k_2}{(2\pi)^d} \phi_s(-k_1-k_2) \psi_{a,s} (k_1)F_{ab}(k_1,k_2) \psi_{b,s}(k_2)
\ee
where $k := -k_1-k_2$, the solid vertex denotes an insertion of the U(1) breaking operator $h_2$, and
\be
	F(k_1,k_2) = -\frac{\lambda_1h_2}{2}\int_f d^dp \Bigg[C [p*(k-p) + k_1*k_2] \tr[CG(p)CG^T(k-p)]
\ee
\[
	-4k_2*(k-p) CG(p)CG^T(k-p) C\Bigg]
\]
Keeping at most two powers of slow momenta $k$, and dropping terms that vanish upon integration, we find
 \be \label{eq:oneloop:F}
 	F(k_1,k_2)  
	= -\lambda_1h_2 C[3k_1*k_2 + 2k_2*k_2] \frac{2}{(4\pi)^{d/2}\Gamma(d/2)} \Lambda^{d-2}\delta l 
\ee

\vspace{10mm}

\begin{figure}[ht]
\begin{center}
\includegraphics[width=0.3\textwidth]{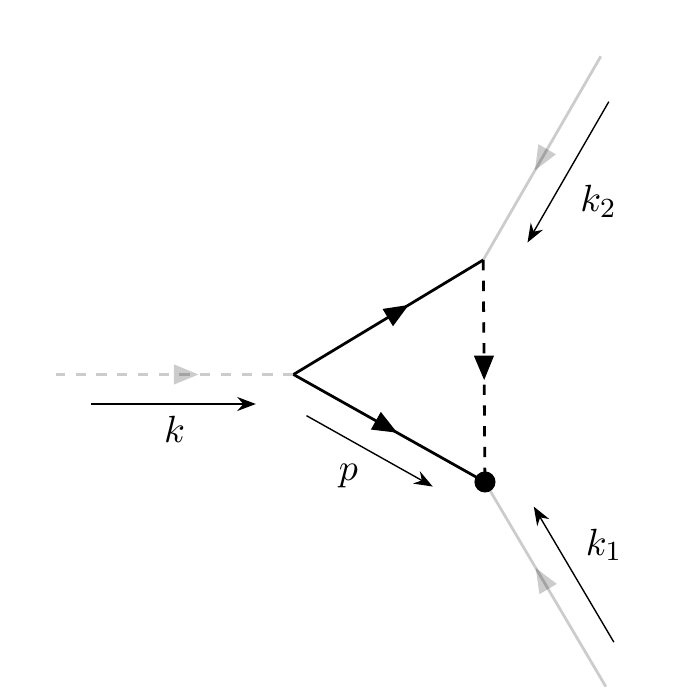}
\caption{Second diagram renormalizing $h_3$ and $h_4$ in Wilson RG}
\label{oneloop:triangle}
\end{center}
\end{figure}

The second diagram renormalizing  $h_3$ and $h_4$ is shown in Figure \ref{oneloop:triangle}, and equals
\be
\int \frac{d^dk_1}{(2\pi)^d} \frac{d^2k_2}{(2\pi)^d} \phi_s(-k_1-k_2) \psi_{a,s} (k_1)G_{ab}(k_1,k_2) \psi_{b,s}(k_2)
\ee
where $k:=- k_1 - k_2$, and the solid vertex denotes the insertion of the U(1) breaking operators proportional to $h_3$ and $h_4$, and 
\be
	G(k_1,k_2) = 
	-4\lambda_1^2 \int_f \frac{d^dp}{(2\pi)^d} D(p) CG(p-k_1) CG^T(-p-k_2) 
\ee
\[
	\times
	 \left[ h_3 k_2*(-p-k_2) + 2h_4(p+k_2)*(p+k_2)\right]
\]
Again, we drop terms proportional to $p*p$ and $k*p$, since they will integrate to zero. The result is, to quadratic order in the slow momenta $k$, 
\be
	G(k_1,k_2) 
	=-4(2h_4-h_3)k_2*k_2\lambda_1^2 C \frac{2}{(4\pi)^{d/2}\Gamma(d/2)} \Lambda^{-\epsilon}\delta l 
\ee

Adding this result to (\ref{eq:oneloop:F}), we find the following renormalization constants:

\be
	Z_{h_3} =  1 -\frac{ 6\lambda_1 h_2 }{h_3} \frac{2\Lambda^{-\epsilon}\delta l}{(4\pi)^{d/2}\Gamma(d/2)}
\ee 
and
\be
	Z_{h_4} =  1+ \left[ 2\lambda_1h_2 + 4(2h_4-h_3)\lambda_1^2\right]\frac{2\Lambda^{-\epsilon}\delta l}{(4\pi)^{d/2}\Gamma(d/2) h_4}
\ee

The factors of $\Lambda^2$ were removed by redefining the couplings constants to be dimensionless from the start of the calculation. For all remaining diagrams, we cite the calculations of [\onlinecite{Zerf:2016fti}], since these do not receive corrections from the U(1) breaking terms or the fermion mass to this order.  As a result, the beta functions for $\lambda_1$ and $\lambda_2$ are unchanged, and we can use the critical value $\lambda_1^2$ from [\onlinecite{Zerf:2016fti}]:
\be \label{eq:oneloop:lambda}
	\frac{\lambda_{1,*}^2 }{(4\pi)^2}= \frac{\epsilon}{12} + \fO(\epsilon^2)
\ee

\subsubsection{Renormalization Constants at $\fO(\epsilon)$}

To determine the value of these renormalization constants to $\fO(\epsilon$), 
we replace $\lambda_1^2$ in these expressions with $\lambda_{1,*}$ in (\ref{eq:oneloop:lambda}). Any corrections from U(1) breaking operators or the fermion mass will be higher order in the parameters $\{h_i,M\}$. We find, to $\fO(\epsilon)$, the following renormalization coefficients:
\begin{align} 
Z_\psi = & 1 + \frac{\epsilon}{3}\delta l  \label{eq:z1} \\
Z_M =  & 1 - \frac{2\epsilon}{3 }\delta l  \\
Z_{h_2} = & 1 \\
Z_{h_3} =  &1 -\frac{6h_2  \delta l}{h_3\sqrt{3}(4\pi)}\sqrt{\epsilon}\\
Z_{h_4} =  &1+ 2\delta l \left[\frac{h_2\sqrt{\epsilon}}{h_44\pi \sqrt{3}}  + \frac{\epsilon(2h_4-h_3)}{3h_4}\right]  \label{eq:z2}\\
\end{align}

\subsection{U(1) Breaking Operators with Repulsive Interactions} \label{app:u1:repulsive}

We now calculate the renormalization constants for the theory (\ref{fl1}). According (\ref{eq:oneloop:h1}), to determine the $h_1$ beta function, we only have to calculate $Z_\psi$ and $Z_{h_1}$. Since there is no one loop diagram renormalizing $h_1$, calculating the fermion propagator will be sufficient. Note that we are using the same symbol $Z_\psi$ for the renormalization constant in both (\ref{fl1}) and (\ref{fl2}), even though they are different quantities. Using the modified $\epsilon$-expansion, the fermion and boson propagators are
\be
	G(p) = \frac{i\stp}{p^2} \hspace{10mm} D(p) = \frac{1}{p^2}
\ee
The fermion mass is set to zero since time reversal symmetry is present at the transition $g_{\text{c},1}$.  We use solid lines (with an arrow indicating the direction of charge) to represent the fermion propagators, and dashed lines to represent the boson propagators. As before, we include the operators of the external legs in the definitions of our Feynman diagrams.

\subsubsection{Fermion Propagator} 
\begin{figure}[ht]
\begin{center}
\includegraphics[width=0.3\textwidth]{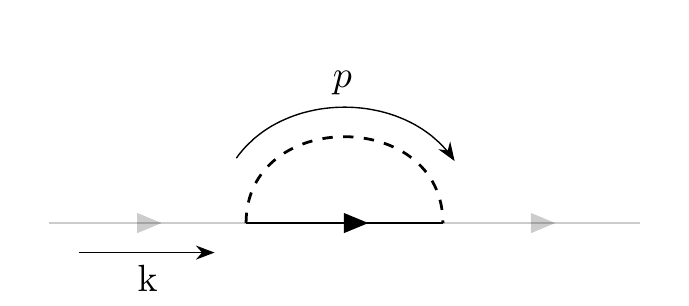}
\caption{Fermion self energy in Wilson RG for $g<0$}
\label{diag:gn:1}
\end{center}
\end{figure}

The single one loop diagram that renormalizes the fermion propagator to $\fO(h_1)$ is shown in Figure \ref{diag:gn:1}. It equals 
\be
	= \int\frac{d^dk}{(2\pi)^d} \bar\psi_s(k) \Sigma_\psi(k) \psi_s(k)
\ee
where 
\be \label{eqA}
	\Sigma_\psi(k) = \eta_1^2 \int \frac{d^dp}{(2\pi)^d} D(p) G(k+p) 
\ee
We expand $\Sigma_\psi(k)$ in powers of $k$, and extract the linear piece to determine 
\be
	Z_\psi = 1 + \left[ 1 - \frac{2}{d}\right] \frac{2\eta_1^2 \Lambda^{-\epsilon} }{(4\pi)^{d/2}\Gamma(d/2)}
\ee 
where we've replaced $\stk$ with $\stk^\dag$, since the difference renormalizes the operator $\bar\psi k_3\psi$, which doesn't enter into the modified $\epsilon$-expansion. 

Since the beta functions for $\eta_1,\eta_2$ receive no $\fO(h_1)$ corrections, we can cite the results of [\onlinecite{Klebanov2016}] that at the critical point $g_{\text{c},1}, \eta_1$ has a value of 
\be
	\frac{\eta_{1,*}^2}{(4\pi)^2} = \frac{\epsilon}{8}  + \fO(\epsilon^2)
\ee
so that to $\fO(\epsilon)$, 
\be \label{eq:z3}
	Z_\psi = 1 + \frac{\eta_1^2}{(4\pi)^2 } = 1 + \frac{\epsilon}{8}
\ee

\section{Superspace Formalism}  \label{app:susy}

In this appendix, we rewrite the Lagrangian density (\ref{fl2}) in superspace notation, at the critical point $g_{\text{c},2}$, where the two U(1) invariant couplings $\lambda_1$ and $\lambda_2$ flow to a common value, $\lambda_*$. We use the results of Section \ref{chapter:u1} to ignore all U(1) breaking operators. This rewriting is most easily done in real time. We introduce a chiral superfield
\be
	\Phi(y) := \phi(y) +\sqrt{2}\theta\psi(y) + \theta^2F(y)
\ee
where $\theta,\bar\theta$ are two-component Grassmann spinors, and $y$ is the (real time) superspace coordinate 
\be
	y^\mu := x^\mu -i\theta\gamma^\mu_R \bar\theta
\ee
By real time, we mean that $x^\mu$ is a  real time coordinate, and the matrices $\gamma^\mu_R = \{-\gamma^0, i\gamma^1, i\gamma^2\}$ satisfy the 2+1 dimensional Minkowski metric:
\be
	\{\gamma^\mu_R, \gamma^\nu_R\} = 2\text{diag}(1,-1,-1)
\ee
 Throughout, we use the following spinor summation convention:
\be
	\theta^\alpha = \epsilon^{\alpha\beta}\theta_\beta \hspace{10mm} \theta_\alpha = \epsilon_{\alpha\beta}\theta^\beta\hspace{10mm}
	\theta^2 = \theta^\alpha\theta_\alpha = 2\theta_2\theta_1
\ee
where 
\be
	\epsilon_{\alpha\beta} := \begin{pmatrix} 0 & -1 \\
	1 & 0 \\
	\end{pmatrix}
	\hspace{15mm}
	\epsilon^{\alpha\beta} := \begin{pmatrix} 0 & 1 \\
	-1 & 0 \\
	\end{pmatrix}
\ee
The Grassmann integration measure is defined as follows:
\be
	d^2\theta = -\frac{1}{4}d\theta^\alpha d\theta^\beta\epsilon_{\alpha\beta} 
	\hspace{10mm} \implies \hspace{10mm} \int d^2\theta \theta^2 =1
\ee
By Taylor expanding $\Phi(y)$ in powers of $\theta$, and integrating out the auxiliary field $F$, one can show that 
\be \label{eq:susy:result}
	\fL_{\text{SUSY}} := \fL_{\text{SUSY}}^0 + \delta \fL_{\text{SUSY}}
	= \partial_\mu \phi^* \partial_\nu\eta^{\mu\nu} \phi + i \bar\psi \gamma^{\mu }_R\partial_\mu \psi
	- \lambda^2|\phi|^4 - \lambda\left(\phi\psi^T C \psi + \text{h.c.}\right)
\ee
where
\be
	\fL_{\text{SUSY}}^0 = \int d^2\theta d^2\bar\theta \Phi^\dag \Phi 
\ee
and
\be
	 \delta \fL_{\text{SUSY}} =  \int d^2\theta W(\Phi) + \int d^2\bar\theta W(\Phi^\dag) \hspace{10mm} W(\Phi) := \frac{\lambda}{3}\Phi^3
\ee
Equation (\ref{eq:susy:result}) is exactly the real time version of (\ref{fl2}), at the critical point $\lambda_1=\lambda_2=\lambda \equiv \lambda_*$. 

\section{two loop Calculation of Fermion Mass Beta Function} \label{app:twoloop}
As a check of (\ref{3:1}), we explicitly calculate the fermion mass beta function in renormalized perturbation theory at two loops in the modified $\epsilon$-expansion. In the $\overline{MS}$ scheme, we find the following renormalization constants:
\be
	Z_\psi = 1 -\frac{4\lambda_1^2}{(4\pi)^2} - \frac{16\lambda_1^4}{(4\pi)^4\epsilon^2} + \frac{8\lambda_1^4}{(4\pi)^4 \epsilon}
\ee
\be
	Z_M = 1 + \frac{8\lambda_1^2}{(4\pi)^2\epsilon}+ \frac{80\lambda_1^4}{(4\pi)^4\epsilon^2}  - \frac{40\lambda_1^4}{(4\pi)^4\epsilon^2}
\ee
The beta function is
\be
	\beta_M = M\left[1 +\gamma_\psi - \gamma_M\right]
	 = M -   \frac{12\lambda_1^2M}{(4\pi)^2} + \frac{96\lambda_1^4M}{(4\pi)^4}
\ee
Using the critical value of $\lambda_1^2$ found in [\onlinecite{Zerf2016}],
\be
	\frac{\lambda_{1,*}^2}{(4\pi)^2} = \frac{\epsilon}{12} + \frac{\epsilon^2}{36}
\ee
the beta function equals 
\be
	\beta_M(\lambda_{1,*}) = \left[1 - 12\left[ \frac{\epsilon}{12} + \frac{\epsilon^2}{36}\right] + 96\frac{\epsilon^2}{144}\right]M
	= \left[1 - \epsilon  +\frac{\epsilon^2}{3}\right]M
\ee
which agrees with the relations (\ref{3:1}) and (\ref{eq:omega}) to $\fO(\epsilon^2)$.

\section{Consequence of a Relevant Fermion Mass Operator} \label{app:largem}
A relevant fermion mass implies that at energy scales $\Lambda \ll M$, the critical point will be described by a purely bosonic theory, obtained by integrating out the fermionic modes completely. To perform this integration explicitly, we use a Hubbard-Stratonovich transformation to replace \emph{all} of the four-Fermi interactions in (\ref{eq1:10}) with 
\be \label{eq3:1}
	\fL_{\text{int}} = -m^2|\phi|^2 + (\phi[\rho_1\bar\psi C\bar\psi^T  +\rho_2\partial_r\psi^TC \partial_r\psi] + \text{h.c.})
\ee
where 
\be
	\rho_1 = 4m\sqrt{g}\Lambda_0^{-1} \hspace{10mm} \rho_2 = \frac{m}{2}\frac{\sqrt{g}}{\Lambda_0^3}
\ee
This expression (\ref{eq3:1}) reproduces (\ref{eq1:10}) to $\fO(g)$ when $\phi$ is integrated out. The boson $\phi$ no longer corresponds to the Cooper pair $\phi \sim \psi_1\psi_2$ of (\ref{fl2}); instead, it corresponds to
\be \label{eq3:2}
	\phi \sim \psi_1\psi_2 + \frac{1}{2}\partial_r\psi_1^*\partial_r\psi_2^*.
\ee
We can use (\ref{eq3:2}) to determine how $\phi$ transforms under the exact lattice symmetries (\ref{eq1:6} - \ref{eq1:9}). Explicitly, these transformations are 
\begin{align} 
C: & \hspace{10mm} \phi(x,y) \mapsto \phi^*(x,y)  \label{eq3:6} \\
T: &  \hspace{10mm} \phi(x,y) \mapsto - \phi^*(x,y)  \label{eq3:7}, \hspace{5mm} i \mapsto -i \\
P:  & \hspace{10mm} \phi(x,y) \mapsto \phi^*(-x,y)  \label{eq3:8}\\
R: & \hspace{10mm} \phi(x,y) \mapsto i \phi(-y,x)
 \label{eq3:9}
\end{align}

\begin{figure}[ht]
\begin{center}
\includegraphics[width=0.5\textwidth]{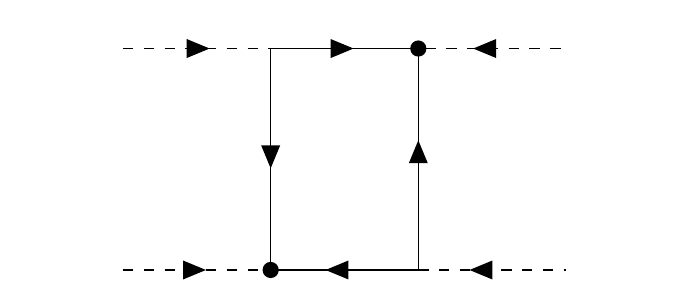}
\caption{Diagram generating $\phi^4 +\text{h.c.}$ when the fermion mass is relevant}
\label{diag10}
\end{center}
\end{figure}

The most noteworthy equation is (\ref{eq3:9}), since it implies that the most relevant U(1) breaking operator allowed by symmetry is $\phi^4 + \phi^{*4}$. To determine the coefficient of this operator, we integrate out the fermions explicitly, using the notation introduced in Chapter \ref{chapter:u1}. The unique one loop diagram generating a $\phi^4$ interaction is shown in Figure \ref{diag10}. 
We are not interested in derivative operators, so we can set all external momenta to zero. The contribution to the operator $\phi^4$ is then equal to 
\be \label{eq:relevant:2}
	= - 8\rho_1^2\rho_2^2 \int \frac{d^3p}{(2\pi)^3} (p*p)^2 \tr [G(p) C G^T(-p) C G(p) C G^T(-p) C]
\ee
where the integral is over all momentum modes up to a cutoff $\Lambda \sim M$. Using $C (\slp^T +M)C = \slp - M$, the trace equals 
\be
	 \tr [G(p) C G^T(-p) C G(p) C G^T(-p) C] 
	 =\frac{2}{(p^2+M^2)^2}
\ee 
Writing  $p*p = p^2\sin^2\theta\cos(2\phi)$ in spherical coordinates, the expression (\ref{eq:relevant:2}) equals
\be
	  - 16\rho_1^2\rho_2^2 \int \frac{d^3p}{(2\pi)^3}  \frac{p^4 \sin^4\theta\cos^2(2\phi)}{(p^2+M^2)^2}
	  \propto \rho_1^2\rho_2^2 M^3
\ee

Therefore, a $\phi^4  + \text{h.c.}$ operator is generated, with coupling constant proportional to
\be
	\rho_1^2\rho_2^2M^3 \propto  \Lambda_0^{-1} \left(\frac{m}{\Lambda_0}\right)^4 g^2\left(\frac{M}{\Lambda_0}\right)^3 
\ee

Since our original assumption was that the fermion mass is small compared to the bare cutoff, we see that the coefficient of $\phi^4$ is highly suppressed. Therefore, the low energy theory near the critical point $g_{\text{c},2}$ has the following structure
\be \label{eq3:12}
	\fL = |\partial_\mu \phi|^2 + m^2|\phi|^2 +  \rho |\phi|^4 + \tilde \rho (\phi_4 + \phi_4^*) \hspace{10mm}  \tilde \rho \ll \rho
\ee

\end{document}